\documentclass[lettersize,journal]{IEEEtran}
\usepackage{amsmath,amsfonts}
\usepackage{algorithmic}
\usepackage{algorithm}
\usepackage{array}
\usepackage[caption=false,font=normalsize,labelfont=sf,textfont=sf]{subfig}
\usepackage{textcomp}
\usepackage{stfloats}
\usepackage{url}
\usepackage{verbatim}
\usepackage{graphicx}
\usepackage{cite}
\usepackage{color}
\usepackage{soul}
\usepackage{colortbl}
\usepackage{algorithm}
\usepackage[table]{xcolor}

\begin{document}

\title{Learning Task-Specific Sampling Strategy for Sparse-View CT Reconstruction}
\author{Liutao Yang, Jiahao Huang, Yingying Fang, Angelica I Aviles-Rivero, Carola-Bibiane Schönlieb, \\Daoqiang Zhang \IEEEmembership{Senior Member, IEEE}, Guang Yang \IEEEmembership{Senior Member, IEEE}
\thanks{
This work was supported in part by the National Natural Science 
Foundation of China (Nos. 62136004, 62276130), National 
Key R\&D Program of China (Grant No. 2023YFF1204803), the Key Research and Development Plan of 
Jiangsu Province (No. BE2022842)}
\thanks{
L. Yang, and D. Zhang are with the College of Computer Science and Technology, Nanjing University of Aeronautics and Astronautics, China. J. Huang and G. Yang are with the Bioengineering Department and Imperial-X, Imperial College London, UK. L. Yang, J. Huang, Y. Fang and G. Yang are also with National Heart and Lung Institute, Imperial College London, UK. J. Huang and G. Yang are also with National Heart Cardiovascular Research Centre, Royal Brompton Hospital, London, UK; G. Yang is also with School of Biomedical Engineering \& Imaging Sciences, King's College London, UK. A. Aviles-Rivero and CB. Schönlieb are with the Department of Applied Mathematics and Theoretical Physics, University of Cambridge, UK. The corresponding author is Daoqiang~Zhang. (email: dqzhang@nuaa.edu.cn.)}}
\maketitle

\begin{abstract}

Sparse-View Computed Tomography (SVCT) offers low-dose and fast imaging but suffers from severe artifacts. Optimizing the sampling strategy is an essential approach to improving the imaging quality of SVCT. However, current methods typically optimize a universal sampling strategy for all types of scans, overlooking the fact that the optimal strategy may vary depending on the specific scanning task, whether it involves particular body scans (e.g., chest CT scans) or downstream clinical applications (e.g., disease diagnosis).
The optimal strategy for one scanning task may not perform as well when applied to other tasks. To address this problem, we propose a deep learning framework that learns task-specific sampling strategies with a multi-task approach to train a unified reconstruction network while tailoring optimal sampling strategies for each individual task.
Thus, a task-specific sampling strategy can be applied for each type of scans to improve the quality of SVCT imaging and further assist in performance of downstream clinical usage. 
 Extensive experiments across different scanning types provide validation for the effectiveness of task-specific sampling strategies in enhancing imaging quality. Experiments involving downstream tasks verify the clinical value of learned sampling strategies, as evidenced by notable improvements in downstream task performance. Furthermore, the utilization of a multi-task framework with a shared reconstruction network facilitates deployment on current imaging devices with switchable task-specific modules, and allows for easily integrate new tasks without retraining the entire model.


\end{abstract}
\begin{IEEEkeywords}
Sparse-View CT, CT reconstruction, Sampling Strategy, Multi-task Learning.
\end{IEEEkeywords}
\section{Introduction}
Computed Tomography (CT) is a noninvasive and highly detailed imaging technique that is widely used in clinical examinations and diagnoses. Large amounts of X-ray projections used during CT examinations raises safety concerns for patients due to the radiation dose. Sparse-View Computed Tomography (SVCT) has advantages in both radiation dose and imaging speed by using a limited number of projections in one scan \cite{fahrig2006dose,wu2021drone}. However, the image quality of SVCT may be significantly degraded due to insufficient sampling of projection data, which is unacceptable in clinical practice.
\begin{figure}
    \centering
    \includegraphics[width=1\linewidth]{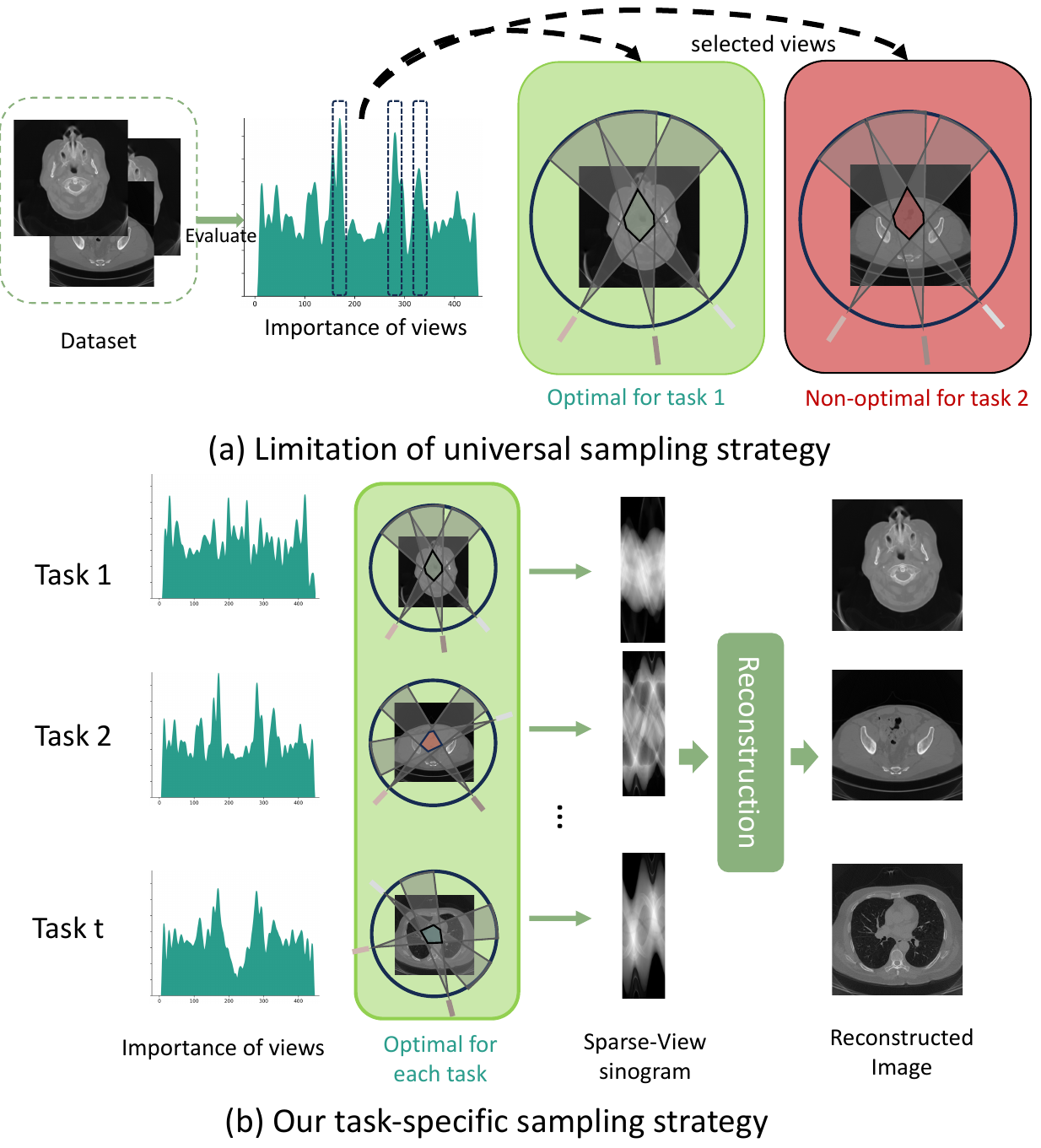}
    \caption{Limitation of universal sampling strategy in SVCT reconstruction \textcolor{cyan}{(a)} and our proposed task-specific sampling strategy \textcolor{cyan}{(b)}. \textcolor{cyan}{(a)} Optimizing a universal sampling strategy for all kinds of CT scans has limitations in finding the optimal strategy for each task. The learned sampling strategy becomes non-optimal when the scanning type changes, thereby undermining the image reconstruction performance. \textcolor{cyan}{(b)} To address this challenge, our proposed task-specific sampling strategy defines different kinds of CT scans as distinct tasks. This approach optimizes strategies for each task within a multi-task framework, aiming to achieve optimal strategies tailored for all types of CT scans. }
    \label{motivation}
\end{figure}

Recently, deep learning reconstruction methods have shown promising performance for sparse-view CT imaging. 
These methods can be categorized into two main approaches. A set of techniques including \cite{han2018framing,kang2017deep,zang2018super,zhao2018sparse} , train an image model to remove artifacts from reconstructed SVCT images. Whilst another body of literature  \cite{he2018optimizing,chen2018learn,xia2023transformer} , unfold the iterative reconstruction process into a deep network and learn the reconstruction parameters through end-to-end training.
In addition to image reconstruction, sampling strategies also play an important role in SVCT imaging.  
Since SVCT uses fewer projection views to reduce the radiation dose, selecting views that contain more information could help reconstruct images with higher quality. Especially in applications such as C-arm CT or dental CT, where scanning time is predominantly determined by the slower flat-panel detector rather than the gantry's mechanical speed, non-uniform sampling emerges as a promising solution for expediting the scanning process\cite{pan2009commercial}. Early works \cite{nagy2005reconstruction,varga2010direction,varga2010projection} have verified the impact of view selection on image quality through experiments or simulations. Some studies have also attempted to optimize sampling strategies using Observer Models \cite{bouhaouel2020task,fischer2016object}.
\begin{figure*}
    \centering
    \includegraphics[width=1\linewidth]{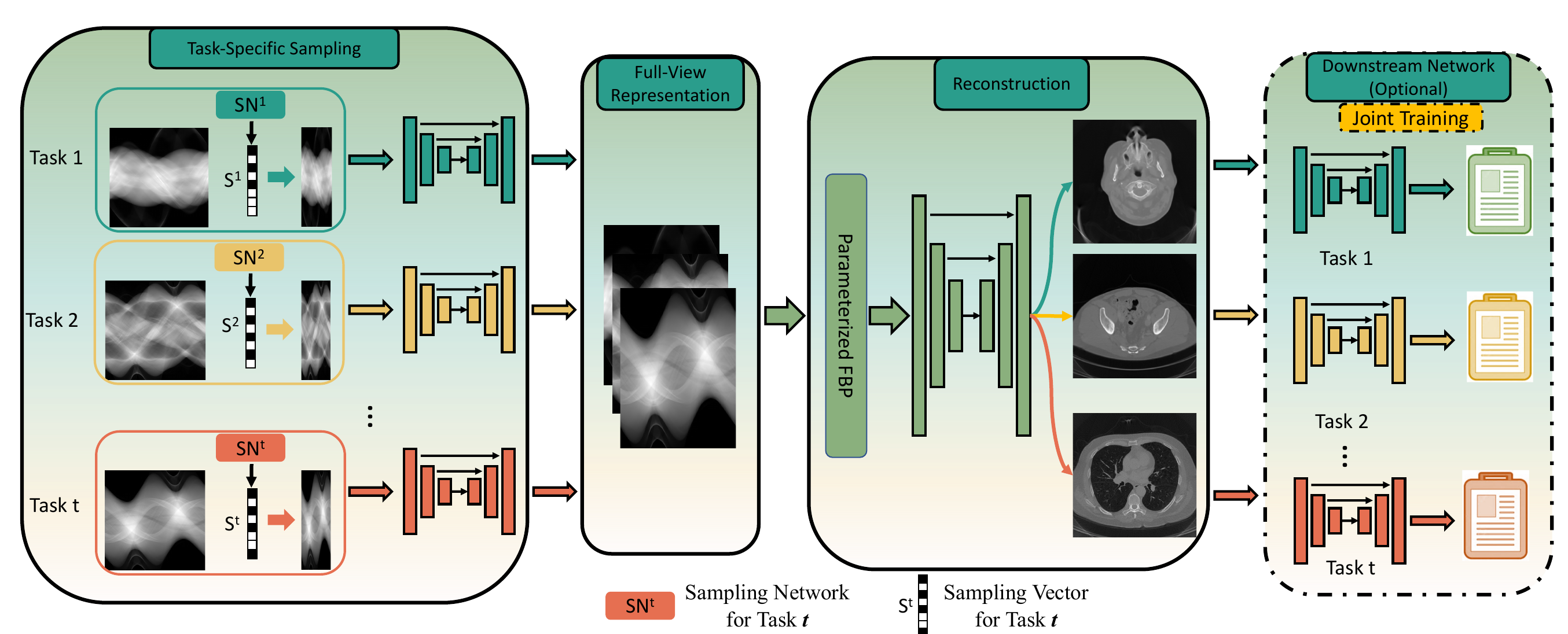}
    \caption{The framework of proposed method. The \textit{task-specific sampling} learns the optimal sampling strategy for each scanning task, while the \textit{reconstruction network} reconstructs high-quality images from the undersampled projections generated by the sampling strategy. The task-specific sampling strategy learning is performed through a multi-task framework. Undersampled projections are completed into full-view using the sinogram complement network for the following reconstruction. If the task branch contains a downstream-task, an downstream-task network would be trained jointly with reconstruction network. }
    \label{framework}
\end{figure*}

More recently, several studies \cite{shen2020learning,wang2022active,yang2022sparse} have explored using deep learning for sampling strategies optimization. These methods usually optimize a universal sampling strategy for all type of scanning tasks. However, different types of scans are always conducted for different goal of examination in clinical practice. Adopting a universal sampling strategy would constrain the capability to capture a tailored sampling strategy for specific scanning scenarios. For instance, the distinction between CT scans of the chest and head, each centering on disparate body regions and organs, might require different sampling strategies for better image restoring. As shown in Fig. \ref{motivation}, the selected views in head scans emphasizes the center area of the head, whereas in chest scans, it highlights the edge area. Thus, optimizing a task-specific sampling strategies has the potential to heighten performance within each task domain by applying diverse configurations aligning with distinct clinical objectives.  Moreover, the performance of reconstructed images on downstream clinical task is also crucial for reconstruction, especially for instances where clinics seek to emphasize scanning for precise disease diagnoses.  

To this end, we propose a task-specific deep learning framework that is designed to learn optimal sampling strategies for various tasks. This framework aims for each task — whether it involves specific body scans (e.g., chest CT scans) or downstream clinical applications (e.g., disease diagnosis) — to achieve outstanding imaging quality or enhance performance in clinical tasks using a limited number of projection views.  This framework trains the reconstruction network in a multi-task manner, with each task corresponding to a specific type, and seeks to achieve high-quality imaging in a small number of views. Unlike previous methods that learn a universal sampling trategy, our approach is tailored. We consider the unique characteristics of each task. This allows us to develop a task-specific sampling strategy. Our method focuses on acquiring the most informative views for each respective task. Consequently, this leads to a significant improvement in image quality when compared to existing methods. Thus, we only need to train the task-specific module and fix parameters in reconstruction network when adding new tasks, which offers a great extensibility for deployment on imaging devices.

Specifically, our proposed framework consists of three components. The first is the sampling network, which is based on a deep neural network to predict the probability distribution of view importance for each scanning task. The second component is the reconstruction network. For all sampling strategy, we train a parameter-shared reconstruction network that reconstructs high-quality images from sparse-view data using optimal sampling strategies. The third component is the downstream-task network, which is a clinical-related network and jointly optimized with reconstructions. Since not all reconstructions correspond to downstream-tasks, the downstream-task networks are optional modules and will only be used in the downstream-task related reconstruction.

We evaluate our approach on various datasets with and without downstream tasks. Experimental results demonstrate that our task-specific sampling strategy significantly outperforms existing methods, and achieves state-of-the-art performance in terms of image quality and downstream tasks using only
a limited number of projection views. Overall, our proposed framework provides a promising solution to the challenge of improving image quality in SVCT, and has the potential to make a significant impact on clinical practice by reducing radiation dose for patients while maintaining high-quality imaging results. Our contributions are summarized next.
\begin{itemize}
\item We propose a deep learning framework for optimizing task-specific sampling strategy in sparse-view CT imaging. Through multi-task learning, this framework can learn sampling strategy separately and share the same network for image reconstruction.

\item We demonstrate the effectiveness of the proposed method in achieving high-quality image reconstruction with a small number of projection views, outperforming other state-of-the-art methods.

\item We highlight the potential benefits in clinical settings, including improved performance on downstream task and reduced radiation doses for patients. Our framework also offers adaptability to various scanning tasks.

\end{itemize}

\section{Related Works}
\subsection{Deep Learning based CT Reconstruction}

A vast amount of literature has been dedicated to improving CT reconstruction tasks through the application of deep learning techniques \cite{jin2017deep,kang2017deep,hu2020hybrid,lee2017view}.
These methods generally fall into two categories: 1) image domain post-processing, and 2) dual-domain reconstruction. Image domain post-processing methods use low-level vision denoising methods
to train an end-to-end network for artifact removing and denoising. For example, Jin \textit{et al.} \cite{jin2017deep} propose the FBPConvNet, which is a CNN based network using the U-net structure. This method tries to improve the quality of the FBP (Filtered Back Projection) reconstructed images using CNNs. 
Han \textit{et al.} \cite{han2018framing} focus on recover high-frequency edges in sparse view CT and use wavelet transform to effectively remove artifacts in deep learning frameworks. Kang \textit{et al.} \cite{kang2017deep} use the directional Wavelets transform in CNNs to denoise in low dose CT. Furthermore, Chen \textit{et al.} \cite{chen2017low} proposed a residual encoder-decoder CNN (REDCNN) for low dose CT. Dual-domain reconstruction methods utilize information from both sinogram and image domains to avoid over-smooth caused by CNN structure.
 Hu \textit{et al.} \cite{hu2020hybrid} introduced HDNet, a hybrid-domain neural network. This network simplifies the SVCT reconstruction challenge into more manageable stages. 
 
 Yang \textit{et al.} \cite{yang2022inner} developed the Sinogram Inner-Structure Transformer. It harnesses sinogram domain structures for LDCT noise reduction. AUTOMAP \cite{zhu2018image}  utilizes a convolutional network to map measurement data to images. Similarly, IRadonMap \cite{he2020radon} converts CT data into image with a parameter learnable inverse Radon transform.

\subsection{Sampling View Selection}

Since SVCT uses fewer projection views to reduce the radiation dose, selecting views that contain more information could help reconstruct images with higher quality. Early works \cite{nagy2005reconstruction,varga2010direction,varga2010projection} have verified the impact of view selection on image quality through experiments or simulations, and some have attempted to optimize sampling strategies using Observer Models \cite{bouhaouel2020task,fischer2016object}.
More recently, several studies have explored the use of deep learning for sampling strategies. For instance, Shen \textit{et al.} \cite{shen2020learning} and Wang \textit{et al.} \cite{wang2022active} incorporate active sinogram sampling into the reconstruction framework. Yang \textit{et al.} \cite{yang2022sparse} propose PLANet, which learns a distribution of view importance to generate the sampling strategy.

\subsection{Task-Specific Reconstruction}
Acquiring a medical image is not the final goal in many clinical applications. Medical images are usually analyzed with post-processing steps to produce clinically relevant parameters \cite{caballero2014application}. Thus, task-specific reconstruction connects both reconstruction and downstream analysis task into one end-to-end training and making reconstruction results adoptive to downstream tasks. Wu \textit{et al.} \cite{wu2018end} propose
a two-step network for lung nodules detection in sparse-view CT scans. The reconstruction network and the CNN detector are trained sequentially first, then followed by one epoch of end-to-end fine tuning. Lee \textit{et al.} \cite{lee2019machine} propose a reconstruction free SinoNet to achieve high performance body region identification and intracranial hemorrhage (ICH) detection form the raw data. Similar techniques are adopted in fast MRI reconstruction.
Sun \textit{et al.} \cite{sun2019joint} developed an algorithm for joint reconstruction and segmentation from undersampled k-space data based on patch-based sparse modelling and Gaussian mixture modelling. Deep learning     based methods mostly consider to train a sequential model by stacking a reconstruction model and a task model, which will be optimized on the multi-task loss \cite{wang2021one, huang2019fr}.

\section{Method}
\subsection{Problem Formulation}

  Let $y$ represent the full-view measured projections. The CT reconstruction problem can then be formulated as the task of estimating the image $x$ from $y$:

\begin{equation}
y = Ax
\end{equation}
where, $y\in R^{J}$, $J=V\times D$, $V$ is the number of projections, and $D$ is the number of detectors. $x\in R^{K}$, where $K=W\times H$, and $W$ and $H$ are the width and height of the image. $A\in R^{J\times K}$ is the measurement matrix.

Sparse-view CT reconstruction can be formulated as estimating $x$ from a subset $y_{s}$ of full-view measurement projections. The sampling strategy can then be defined as a binary vector $S$ that selects views from $y$ to generate $y_{s}$.

\subsection{Overview of the Proposed Framework}
As shown in Fig. \ref{framework}, the proposed method consists of three main components: The sampling strategy learning, the reconstruction network and the downstream-task network.

The sampling strategy learning aims to learn the optimal sampling strategy for each specific scanning task, while the reconstruction network is trained to reconstruct high-quality images from the undersampled projections generated by the sampling strategy. The third part is downstream-task network, which conducts clinical-related tasks directly using the reconstruct images. Since not all reconstructions correspond to downstream-tasks, the downstream-task network is an optional module and will only be used in downstream-task related reconstruction.

 For clarity, we use \textbf{\textit{task}}  to indicate each task branch in the multi-task framework which is a CT reconstruction task on different body part or for specific clinical propose; we use \textbf{\textit{downstream-task}} to represent the specific clinical-related task which uses the reconstruction results as input for clinical analysis (e.g., disease diagnosis, organ segmentation or lesion detection). 
\begin{figure*}[t]
\begin{center}
		\includegraphics[width=1\linewidth]{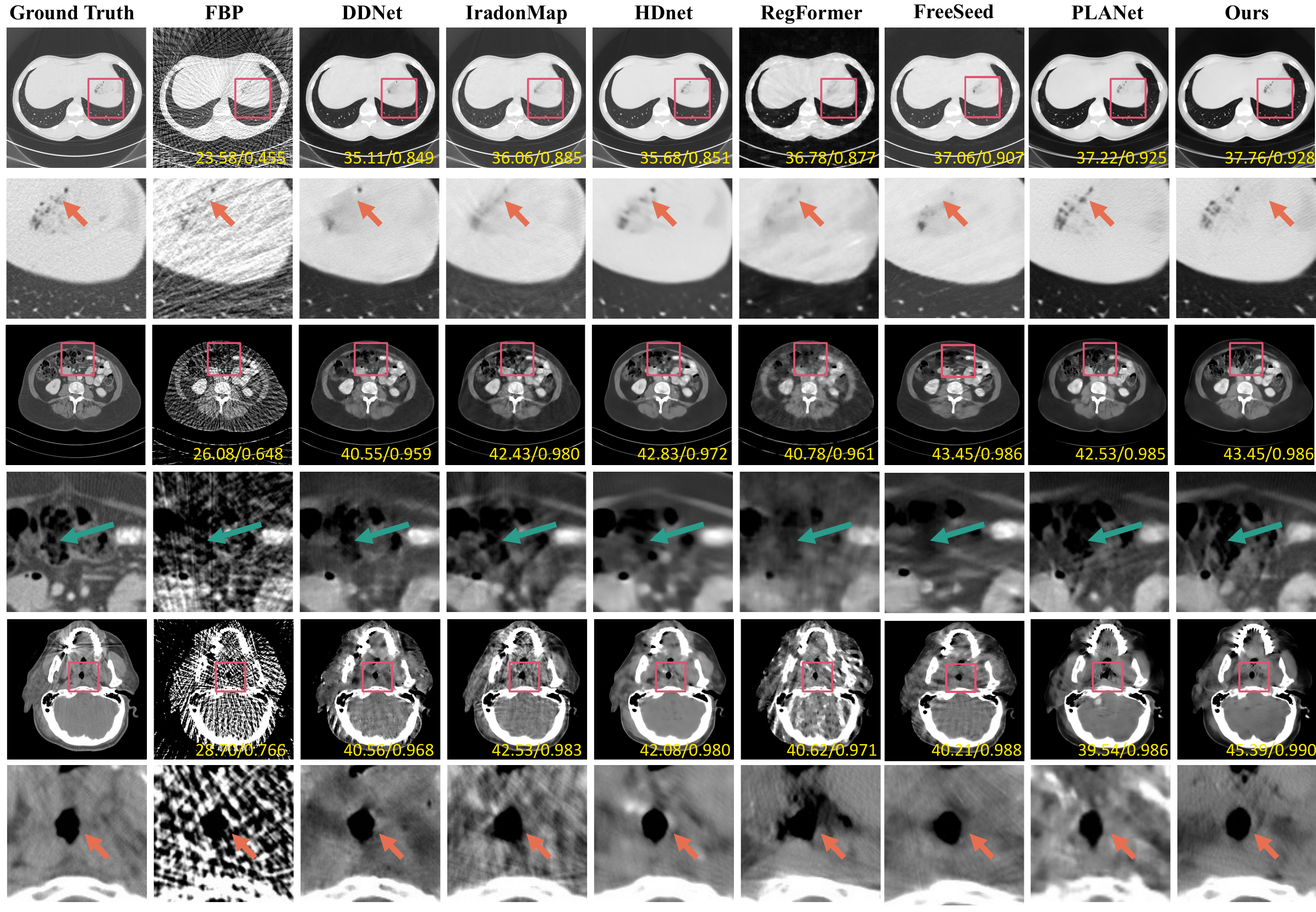}
	\end{center}
	\caption{Visual results for three types of scans: low-dose non-contrast chest scans, contrast-enhanced CT scans of the abdomen, and non-contrast head CT scans (top to bottom). PSNR/SSIM are marked in yellow at the bottom of each image. All images are reconstructed using 60 views, and regions in red boxes are zoomed below each row. As the arrows show, our method has better performance on details recovering for all three types of scan.}
	\label{method_compare}
\end{figure*}

\subsection{Task-Specific Sampling Strategy Learning}
In this section, a multi-task learning framework is used to learn the task-specific sampling strategy. For task branch $t$, the goal is to learn the optimal sampling strategy $S^{t}$, a binary vector that represents the chosen projection views. As shown in Fig. \ref{framework}, for task $t$, we design a network branch, namely the Sampling Network (SN), to learn each sampling strategy separately. Since there is no ground truth for the sampling strategy, it is learned indirectly through the reconstruction training. Specifically, we follow \cite{yang2022sparse} to learn the sampling strategy by embedding a Gumbel-Max \cite{bao2021beit,ramesh2021zero} sampling layer into the network. This assumes that the importance of views can be represented by a learnable distribution $P^{t}$ (representing the importance of each view), and sparse-view sinograms are generated based on this distribution:
 \begin{figure*}[t]
\begin{center}
		\includegraphics[width=0.9\linewidth]{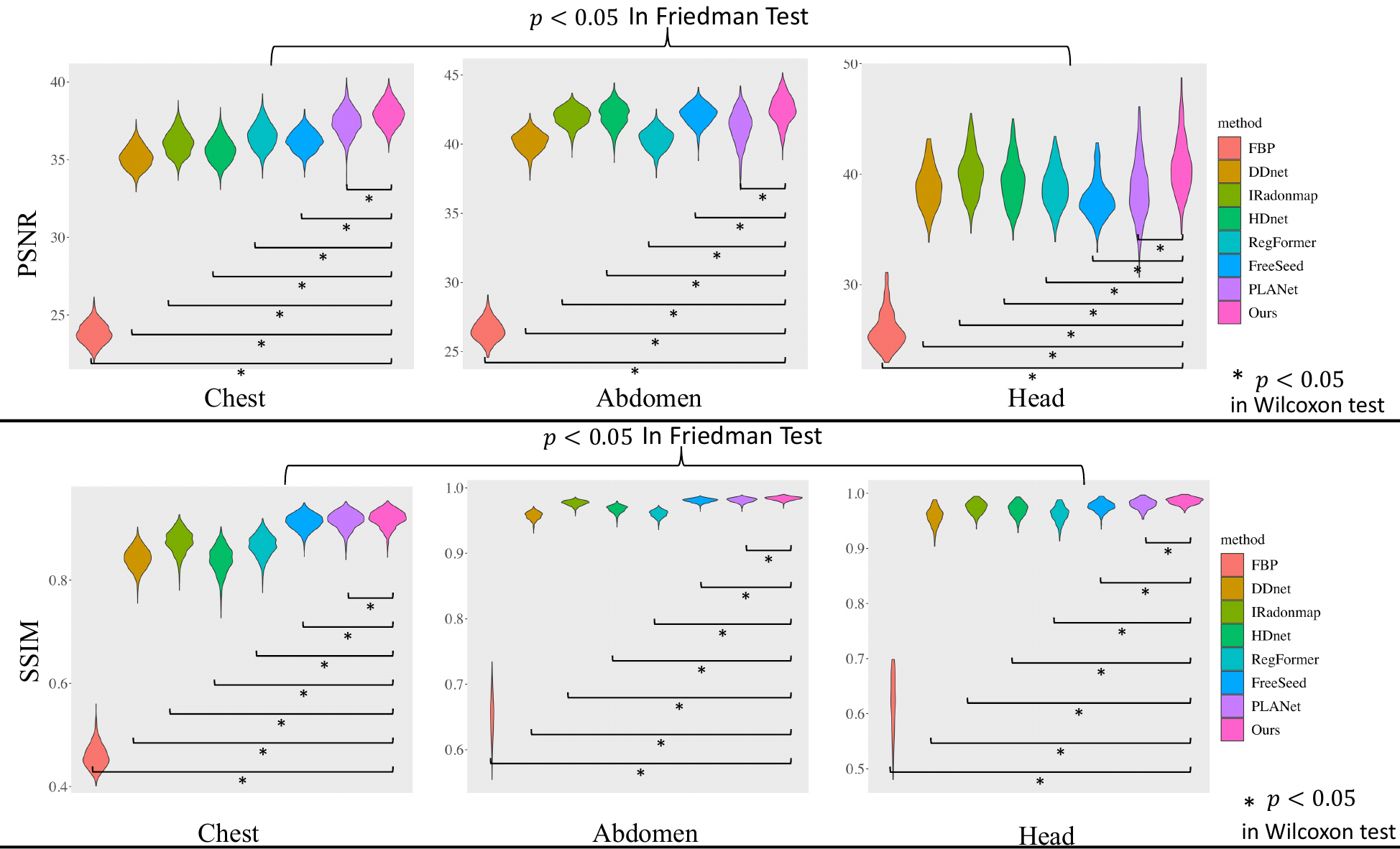}
	\end{center}
	\caption{Violin plots of PSNR (top) and SSIM (bottom) result of sole reconstruction from three tasks. We first conduct the Friedman test to verify if the performances of each method are significant different ($p < 0.05$) over all three tasks. In each task, the Wilcoxon test is conducted between the proposed method and other comparison methods, and $*$ indicates $p < 0.05$.   }
	\label{violin}
\end{figure*}
\begin{equation}
\label{gam}
\begin{array}{ll}
s_{i}^{t}  =\dfrac{\exp \left(\frac{\log \left(P^{t}(\theta_{i})\right)+g_{i}}{\tau}\right)}{\sum_{j=1}^{V} \exp \left(\frac{\log \left(P^{t}(\theta_{i})\right)+g_{j}}{\tau}\right)}, & \\  
\\
g_{i}  =-\log \left(-\log \left(u_{i}\right)\right), &\\
 u_{i} \sim  Uniform (0,1) &    
\end{array}  
\end{equation}
where $s_{i}^{t}$ is the $i_{th}$ element of $S^{t}$, $u_{i}$ is drawn from a uniform distribution, $\theta_{i}$ is the $i_{th}$ projection view, and $\tau$ is a temperature term \cite{yang2022sparse,ramesh2021zero}. By optimizing $P^{t}$ during training, SN can generate sinograms that contain views with higher importance. When training is completed, we can obtain $S^{t}$ as a sample drawn from the distribution $P^{t}$.

\begin{figure}
    \centering
    \includegraphics[width=1\linewidth]{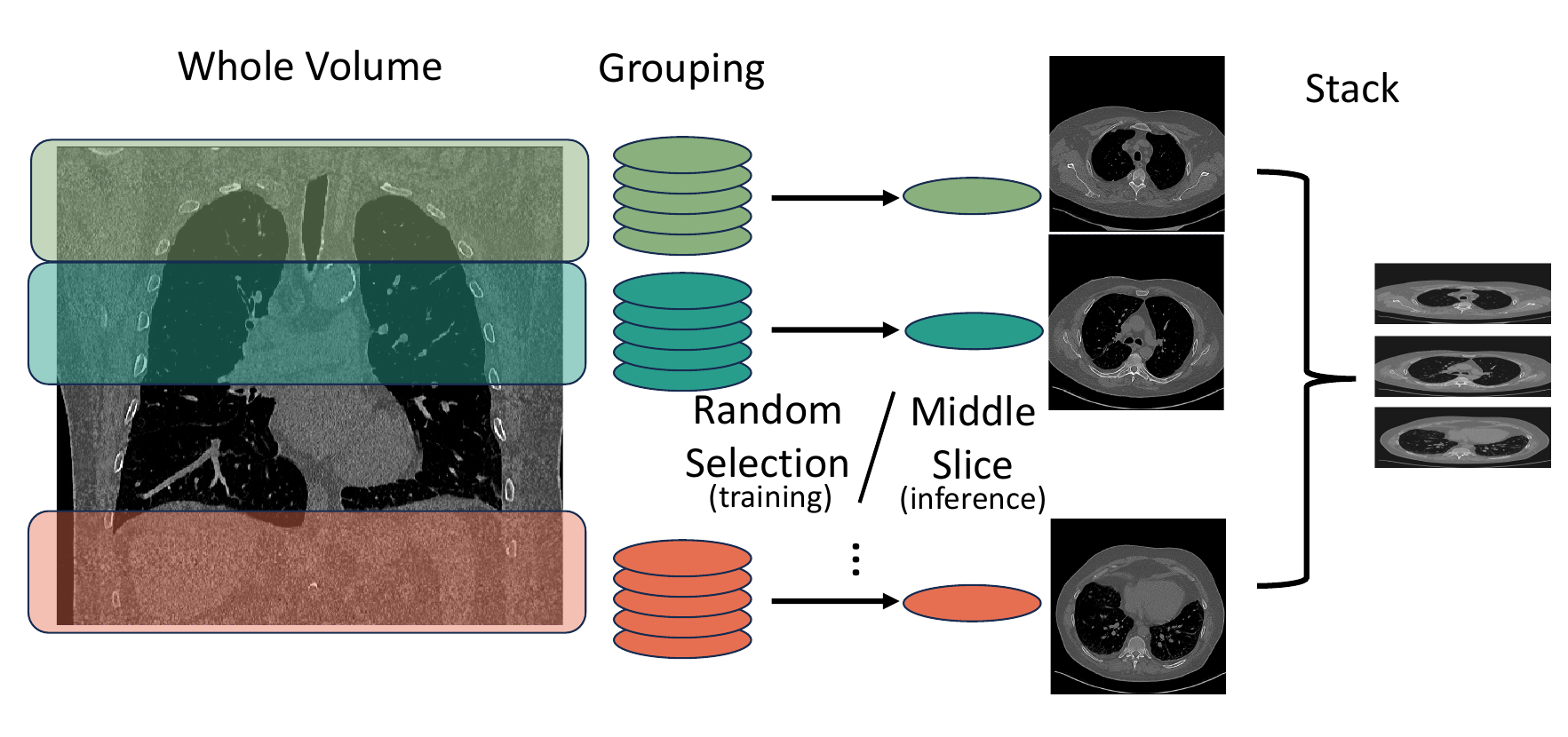}
    \caption{Illustration of slice selection for downstream task. We first split the whole volume of one case into $c$ group. In each group, we select one slice randomly during training as the data augmentation and use the middle slice while testing.}
    \label{slice}
\end{figure}
 In order to deal with the performance degradation of parameter sharing, we attached a sinogram network to interpolate sampled sparse-view sinograms into full-view as the shared representation. Typically, for multi-task model, the design of the shared representation has a significant impact on performance \cite{girdhar2022omnivore}. However, it can be difficult to find a suitable representation that works well for all tasks. In the case of SVCT reconstruction, the full view sinogram serves as a natural shared representation that is suitable for all types of scans. Additionally, ground truth data are available for supervised training of the full-view representation. 
With this design, the parameter-shared reconstruction network can be effectively decoupled from each task and can focus solely on reconstruction.

 \subsection{Parameter-Shared Reconstruction Network}
 As shown in Fig. \ref{framework}, the reconstruction network aims to provide a universal image reconstruction model from sinograms and backpropagate gradients for strategy learning in all task branches. Hence, we first utilize a learning-based FBP \cite{he2020radon} to achieve a coarse reconstruction from input sinograms. With this design, we can fully exploit FBP to backpropagate gradients between two domains and employ learnable parameters to enhance performance. The formulation can be described as follows:
 \begin{equation}
 \label{bp}
    \begin{array}{cc}
    y_{f}(k, j)=\tanh \left(\sum_{i=1}^{P} \eta_{k i} \cdot y(i, j)\right), &\\
    x(n, m)=\left.\sum_{j=1}^{V} \epsilon_{n m j} \cdot y_{f}(i, j)\right|_{i=I N T\left[ \arctan{\big(\frac{a_j}{R-b_j}\big)/\gamma_{d}}      \right]},&\\
    a_j=n \cdot \cos{\theta_{j}}- m \cdot \sin{\theta_{j}},&\\
    b_j=n \cdot \sin{\theta_{j}}+ m \cdot \cos{\theta_{j}}&
     \end{array}
 \end{equation}
where the filtering operation in the FBP (filtered back-projection) algorithm is replaced with a learnable network, $\eta$ represents the parameter of the fully connected layer. The index of detectors in the original sinogram is denoted by $i$, while $k$ represents the index of detectors in the filtered sinogram. The second line describes the parameterized back-projection operation in FBP, with $\epsilon$ representing the learnable parameter. The spatial indices of image $x$ are represented by $n$ and $m$. Each pixel $x(n, m)$ corresponds to a sinusoid in the sinogram data, and $I N T$ refers to the nearest neighbor interpolation method. To calculate the indices of corresponding sinusoid $i,j$ for each $x(n, m)$, the term $I N T\left[ \arctan{(\frac{x_j}{R-y_j})/\gamma_{d}}\right]$ is used, where $\gamma_{d}$ is the interval angle of detectors and $R$ is the distance between the distance between the focal center and detector elements. Then, a U-Net\cite{ronneberger2015u} is used for further adjusting the image quality from the coarse reconstruction.

\subsection{Learning Sampling and Joint Training for Downstream Task}
For downstream-task related reconstruction, a corresponding downstream network $D^{t}$ is attached after image reconstruction if this task is for specific clinical usage. In this paper, the downstream tasks are all defined as the diagnosis for simplification. It can be easily extended into other downstream-tasks such as segmentation or detection by replacing the downstream-task network and loss. To connect slice-level reconstruction network with the case-level downstream-task network, we select $c$ slices from each case and stack them channel-wise before feeding into the downstream-task network:
\begin{equation}
    \text{Out}_{\text{pred}} = D^{t}(stack(x^{t}_{1},x^{t}_{2},...,x^{t}_{c}))
\end{equation}
 Where the $x^t$ is the reconstructed image for task branch $t$. The process of slice selection is shown in \ref{slice}. We first split the whole volume of one case into $c$ group. In each group, we select one slice randomly during training as the data augmentation methods for downstream network. During inference, we use the middle slice in each group for the better stability. To reduce the difficulty of joint optimization, the down stream tasks are fine-tuned in the pretrained reconstruction network. During training, we fix parameters of reconstruction network and update the sampling network and downstream network. Note that the gradient of downstream network $D^{t}$ would be back-propagated into the corresponding sampling strategy learning module to optimize the task-specific sampling $S^{t}$.

\sethlcolor{cyan}	
 
 \begin{table*}[!h]
\begin{center}
\caption{Quantitative results for different methods. Best results are highlighted and second results are \underline{underlined}. $\downarrow$ ($\uparrow$) means the lower (higher) the better. }
\label{t1}
\begin{tabular}{c||cc|cc|cc|cc}
\hline\hline
\rowcolor{gray!20} &\multicolumn{2}{c}{Chest}\vline&\multicolumn{2}{c}{Abdomen}\vline&\multicolumn{2}{c}{Head}\vline&\multicolumn{2}{c}{Average}\\
\rowcolor{gray!20}
Method & PSNR($\uparrow$)&SSIM($\uparrow$)&PSNR($\uparrow$)&SSIM($\uparrow$)&PSNR($\uparrow$)&SSIM($\uparrow$)&PSNR($\uparrow$)&SSIM($\uparrow$)\\
\hline\hline
        FBP  &23.82\scriptsize{$\pm$0.68}&0.459\scriptsize{$\pm$0.0228} &26.55\scriptsize{$\pm$0.77}&0.652\scriptsize{$\pm$0.0314} &25.86\scriptsize{$\pm$1.65}&0.621\scriptsize{$\pm$0.0502} &24.76\scriptsize{$\pm$1.50}&0.527\scriptsize{$\pm$0.095} \\
		DDNet  &35.14\scriptsize{$\pm$0.71}&0.842\scriptsize{$\pm$0.0210} &40.18\scriptsize{$\pm$0.70}&0.957\scriptsize{$\pm$0.0060} &38.82\scriptsize{$\pm$1.86}&0.958\scriptsize{$\pm$0.0160} &36.88\scriptsize{$\pm$2.47}&0.884\scriptsize{$\pm$0.058}	\\
		IradonMap  &36.07\scriptsize{$\pm$0.78}&0.875\scriptsize{$\pm$0.0219} &41.96\scriptsize{$\pm$0.68}&0.977\scriptsize{$\pm$0.0040} &\underline{40.06\scriptsize{$\pm$2.02}}&0.975\scriptsize{$\pm$0.0108} &38.08\scriptsize{$\pm$2.84}&0.912\scriptsize{$\pm$0.052}\\
		
		HDNet&35.61\scriptsize{$\pm$0.78}&0.839\scriptsize{$\pm$0.0256} &42.08\scriptsize{$\pm$0.85}&0.967\scriptsize{$\pm$0.0055} &39.32\scriptsize{$\pm$2.07}&0.971\scriptsize{$\pm$0.0121} &37.77\scriptsize{$\pm$3.08}&0.886\scriptsize{$\pm$0.065}\\
		
		RegFormer&36.53\scriptsize{$\pm$0.80}&0.867\scriptsize{$\pm$0.0211} &40.34\scriptsize{$\pm$0.73}&0.960\scriptsize{$\pm$0.0056} &38.79\scriptsize{$\pm$1.79}&0.962\scriptsize{$\pm$0.0145} &37.80\scriptsize{$\pm$1.94}&0.901\scriptsize{$\pm$0.048}\\
		FreeSeed&36.21\scriptsize{$\pm$0.65}&0.913\scriptsize{$\pm$0.0153}&\underline{42.13\scriptsize{$\pm$0.71}}&\underline{0.981\scriptsize{$\pm$0.0031} }&37.75\scriptsize{$\pm$1.78}&0.976\scriptsize{$\pm$0.0097} &38.05\scriptsize{$\pm$2.77}&0.937\scriptsize{$\pm$0.034} \\
  PLANet&\underline{37.40\scriptsize{$\pm$0.84}}&\underline{0.916\scriptsize{$\pm$0.0165}}&41.25\scriptsize{$\pm$1.18}&\underline{0.981\scriptsize{$\pm$0.0034}} &38.65\scriptsize{$\pm$2.65}&\underline{0.980\scriptsize{$\pm$0.0088}}
        &\underline{38.62\scriptsize{$\pm$2.08}}&\underline{0.940\scriptsize{$\pm$0.034}}\\
		Ours&\  \cellcolor[HTML]{D7FFD7}37.94\scriptsize{$\pm$0.77}&   \cellcolor[HTML]{D7FFD7} 0.919\scriptsize{$\pm$0.0165} &\cellcolor[HTML]{D7FFD7}42.50\scriptsize{$\pm$1.01}&\cellcolor[HTML]{D7FFD7}{0.984}\scriptsize{$\pm$0.0029} &\cellcolor[HTML]{D7FFD7}40.81\scriptsize{$\pm$2.58}&\cellcolor[HTML]{D7FFD7}0.986\scriptsize{$\pm$0.0064} &\cellcolor[HTML]{D7FFD7}39.48\scriptsize{$\pm$2.33}&\cellcolor[HTML]{D7FFD7} 0.943\scriptsize{$\pm$0.034}\\

		\hline
		
	\end{tabular}
\end{center}
\end{table*}
 \subsection{Multi-Task Training and Deployment}
 \label{train}
In typical multi-task learning, a shared representation needs to be learned for the parameter-shared network \cite{sun2020learning,crawshaw2020multi,sun2020adashare}, and the design of sharing representation plays an important role in multi-task training. Rather than learn the representation, a completed full-view sinogram can be a reasonable sharing representation for each sparse-view CT reconstruction task. Moreover, there are labels of full-view sinograms to allow translating each task into the shared representation with supervised training. Thus, it can effectively minimize the performance degradation cost incurred by parameter sharing.

During training, each task branch down-samples a sparse-view sinogram based on the learnable distribution $P^{t}$, which is then completed to full-view sinogram for training the reconstruction network. If this task branch is related with downstream-task, the reconstructed images are further stacked and fed into the following downstream-task network.

\begin{figure}[t]
\begin{center}
		\includegraphics[width=0.85\linewidth]{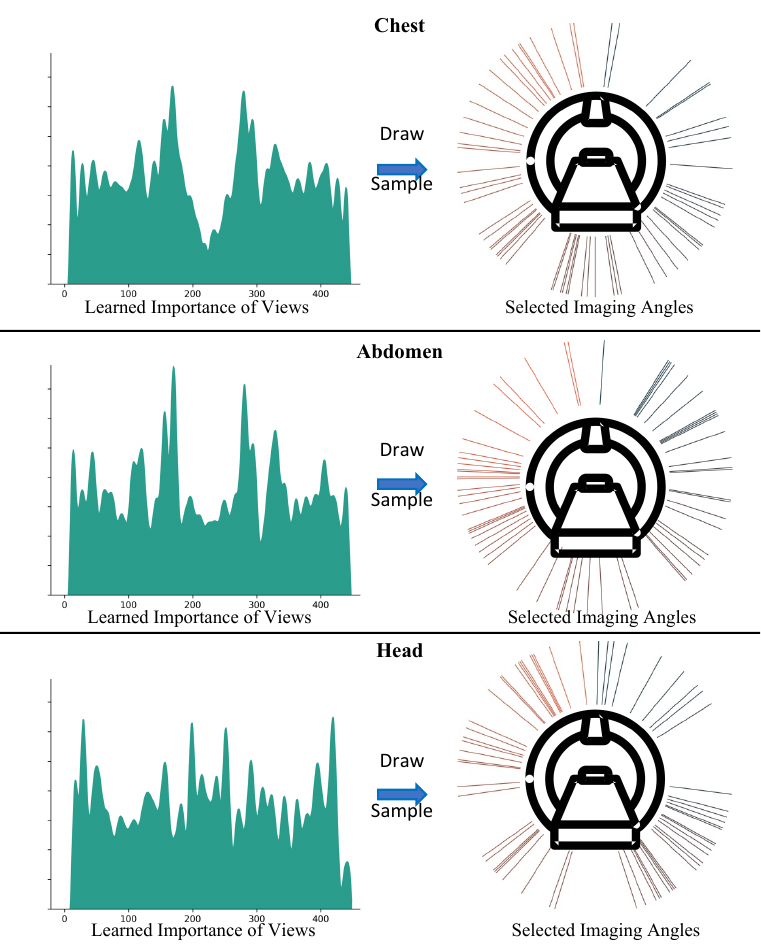}
	\end{center}
	\caption{The learned sampling strategies of three different body parts are depicted. On the left are the learned distributions of view importance, while on the right are the sampling angles drawn from them.}
	\label{views}
\end{figure}

We train all tasks to minimize MSE losses on both sinogram and image domain for reconstruction and the cross entropy loss for classification if downstream task is available. For mini-batch construction, we followed the approach of \cite{girdhar2022omnivore} by constructing mini-batches separately from each dataset and updating network parameters in turn.

After training, we generated a sampling vector $S^{t}$ for each task based on the learned distribution $P^{t}$ and determine the sampling angles for each task to be deployed on the imaging equipment. When conducting a specific type of scanning, we can switch to the corresponding sampling sequence and sinogram network, and reconstruct the images using the parameter-shared reconstruction network. When adding a new task branch, we only need to train the task-specific module and fix parameters in reconstruction network, which offers a great extensibility for deployment on imaging devices. 
\begin{figure}[t!]
\begin{center}
		\includegraphics[width=1\linewidth]{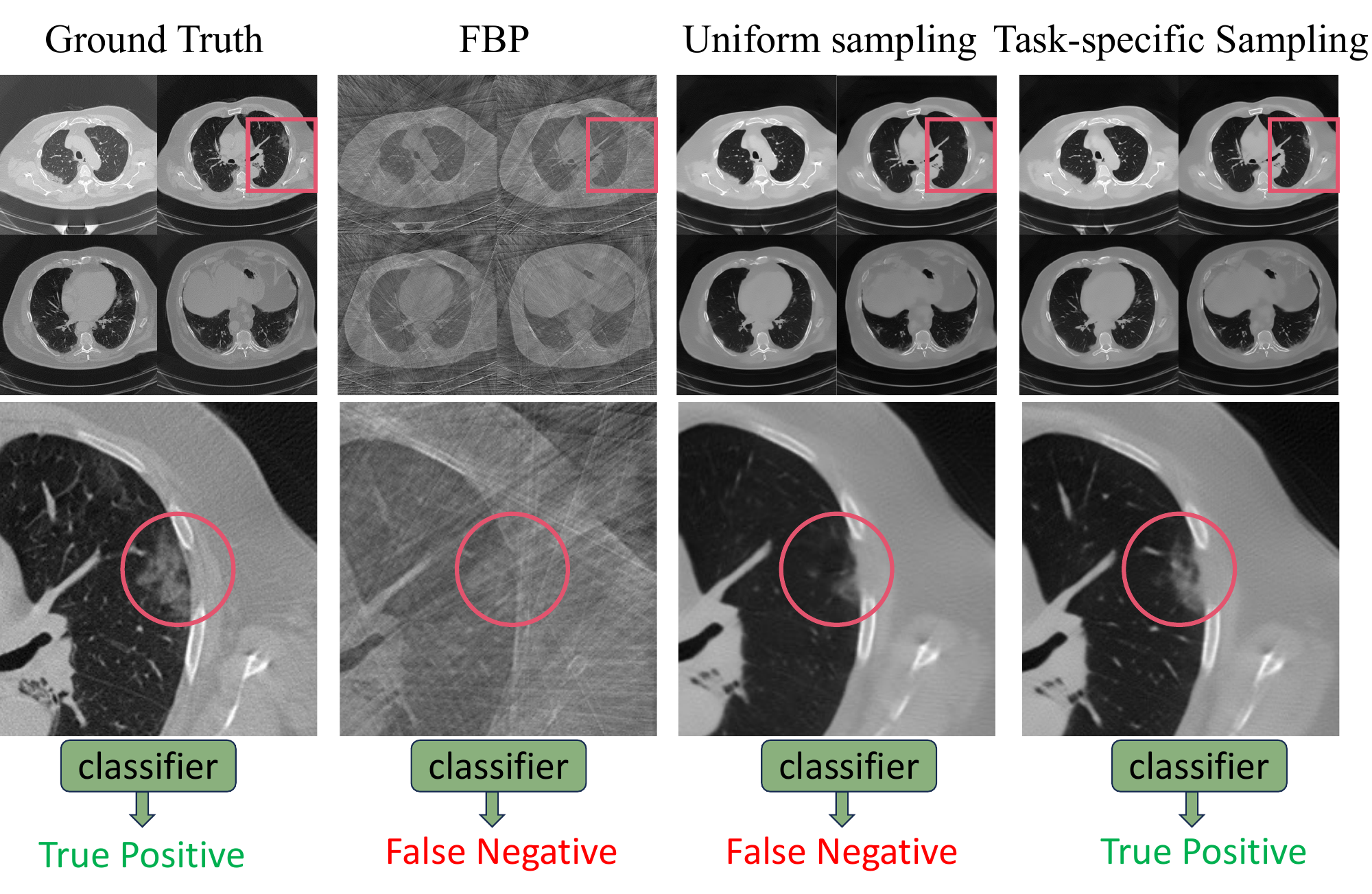}
	\end{center}
	\caption{Examples of compared methods in joint training for downstream-task (Covid-19 diagnosis): 1) Full-view images (GT); 2) Sparse-view images using uniform sampling and reconstructed with FBP (FBP); 3) Sparse-view images using uniform sampling and reconstructed with deep learning network (Uniform Sampling); 4) Sparse-view images using proposed method (Task-Specific Sampling). This example shows that task-specific sampling reconstruction exhibits higher quality in the lesion area compared to uniform sampling and predicts correct result. }
	\label{downstream_f}
\end{figure}

\section{Experiments}
\subsection{Datasets and Setup}
\subsubsection{Datasets for Sole Reconstruction}
For the experiment on sole reconstruction (i.e., framework without downstream-task), each task branch corresponds to a CT scanning type. We carried out experiments on three commonly used types of examinations: non-contrast head CT scans obtained for patients with acute cognitive or motor deficits. low-dose non-contrast chest scans obtained to screen high-risk patients for pulmonary nodules. And contrast-enhanced CT scans of the abdomen obtained to detect metastatic liver lesions. All data were collected from \textit{Low-Dose CT Image and Projection Datasets} \cite{moen2021low} including 40 cases for each type (120 cases in total). We randomly split 
80\% data for training and remain for testing. The full-view sinograms are projected using the ASTRA tomography Toolbox \cite{van2016fast} in a fan-beam CT geometry, and the number of views is 448. The resolution of the reconstructed image is $512 \times 512$. We use both Peak Signal-to-Noise-Ratio(PSNR) and Structure Similarity Index Measure(SSIM) to evaluate performance.

\subsubsection{Datasets for joint Reconstruction with Downstream-task}
 To verify effectiveness of task-specific sampling on downstream-task, we choose two separated COVID-19 related task for the experiments.

\textbf{\textit{COVID-19 Diagnosis}}: We use the \textit{iCTCF dataset} \cite{ning2020open} for this task, which is a patient-centric resource named integrative CT images and clinical features (CFs) for COVID-19 (iCTCF) to archive chest CT images, 130 types of CFs and laboratory-confirmed SARS-CoV-2 clinical status from 1521 patients with or without COVID-19 pneumonia. In our experiments, we only use the CT images for classifying COVID-19 positive or negative. We randomly split 75\% of patients for training and rest for testing. 

\textit{\textbf{Survive/Death Prediction}}: We use the \textit{Saber Dataset \cite{fang2024dynamic}} for performance evaluation. This is an inhouse dataset from 568 patients with COVID-19 pneumonia. In our experiments, we use the CT images for prediction of the survive or death of patients in ten-days after confirmed COVID-19 positive. There are 310 cases of survival and 258 deaths observed within ten days. We randomly split 75\% of patients for training and rest for testing. The full-view sinograms are generated using the same configure as in sole reconstruction. We evaluate the performance of diagnosis tasks using the Accuracy(Acc), Sensitivity(Sens), and Specificity(Spec).

\begin{figure}[t!]
\begin{center}
		\includegraphics[width=0.8\linewidth]{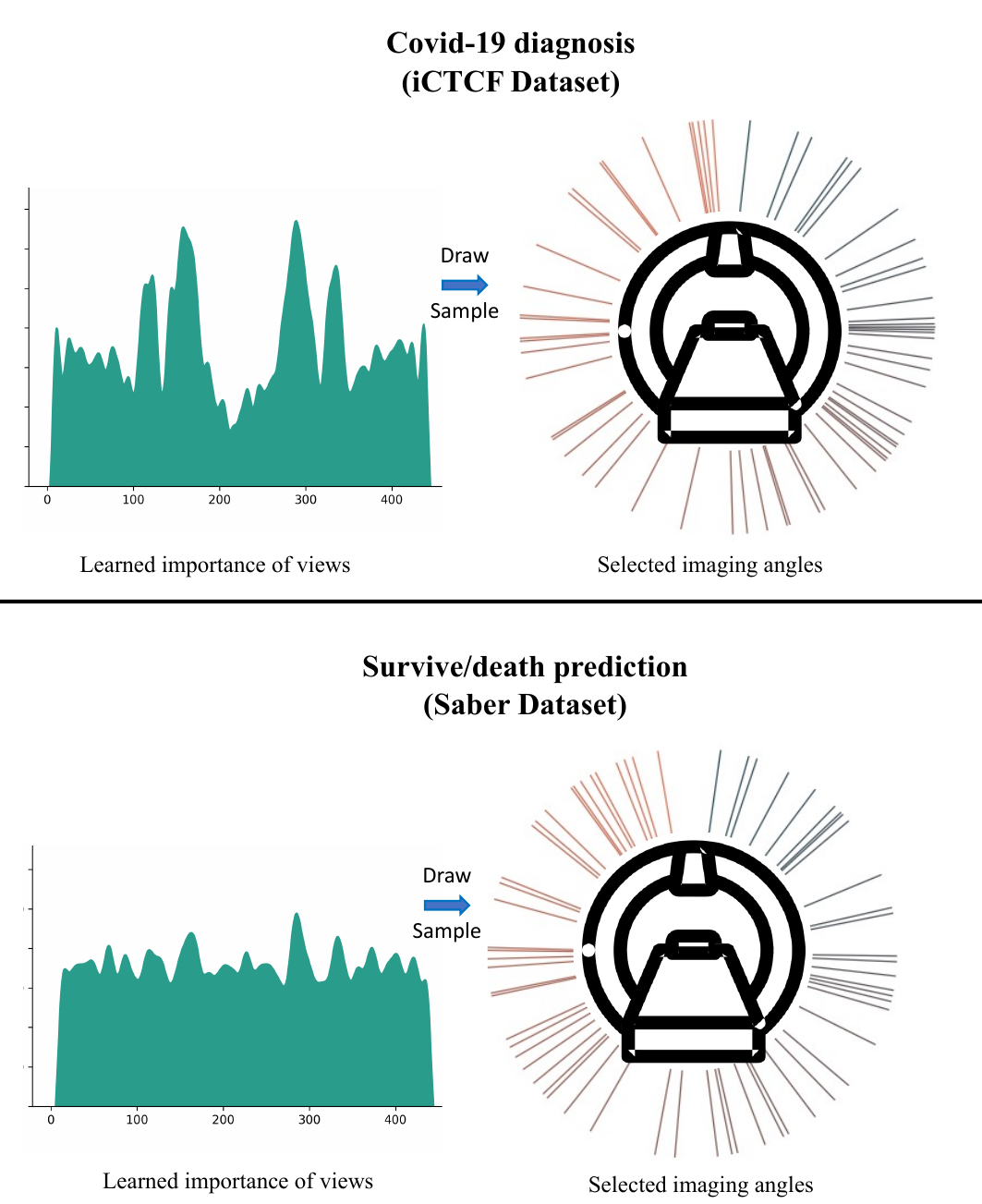}
	\end{center}
	\caption{Learned sampling strategies of two downstream tasks. On the left are the learned distributions of view importance, while on the right are the sampling angles drawn from them.}
	\label{viewsd}
\end{figure}

\begin{table*}[!h]
\begin{center}
\caption{Quantitative results for downstream-task. GT represents results using full-view CT images. Others are results using images that reconstructed from sparse-view sinogram (60 views).Best results are highlighted and second results are \underline{underlined}.}
\label{downstream}
\begin{tabular}{c||ccc|ccc}
\hline\hline
\rowcolor{gray!20}&\multicolumn{3}{c}{Covid-19 diagnosis
}\vline&\multicolumn{3}{c}{Survive/death prediction}\\
\rowcolor{gray!20}&\multicolumn{3}{c}{(iCTCF Dataset)}\vline&\multicolumn{3}{c}{(Saber Dataset)}\\
\rowcolor{gray!20} Method & Acc&Sens&Spec&Acc&Sens&Spec\\
\hline\hline
 GT&\cellcolor[HTML]{D7FFD7}84.27&\cellcolor[HTML]{D7FFD7} 95.75&\cellcolor[HTML]{D7FFD7}68.35  &\cellcolor[HTML]{D7FFD7}66.43&71.79&\cellcolor[HTML]{D7FFD7}60.00 \\ 
 FBP&75.28& 78.19& 56.96  &54.54&\cellcolor[HTML]{D7FFD7}99.99&0.00\\
 Uniform Sampling&77.90&87.77&54.43&61.54 &\underline{92.31}&  24.63\\
 Task-Specific Sampling&\underline{81.27}&\underline{90.43}&\underline{59.49}&\underline{62.24}&70.51&  \underline{52.31} \\

		\hline
		
	\end{tabular}
\end{center}
\end{table*}
\begin{figure}[t]
\begin{center}
		\includegraphics[width=1\linewidth]{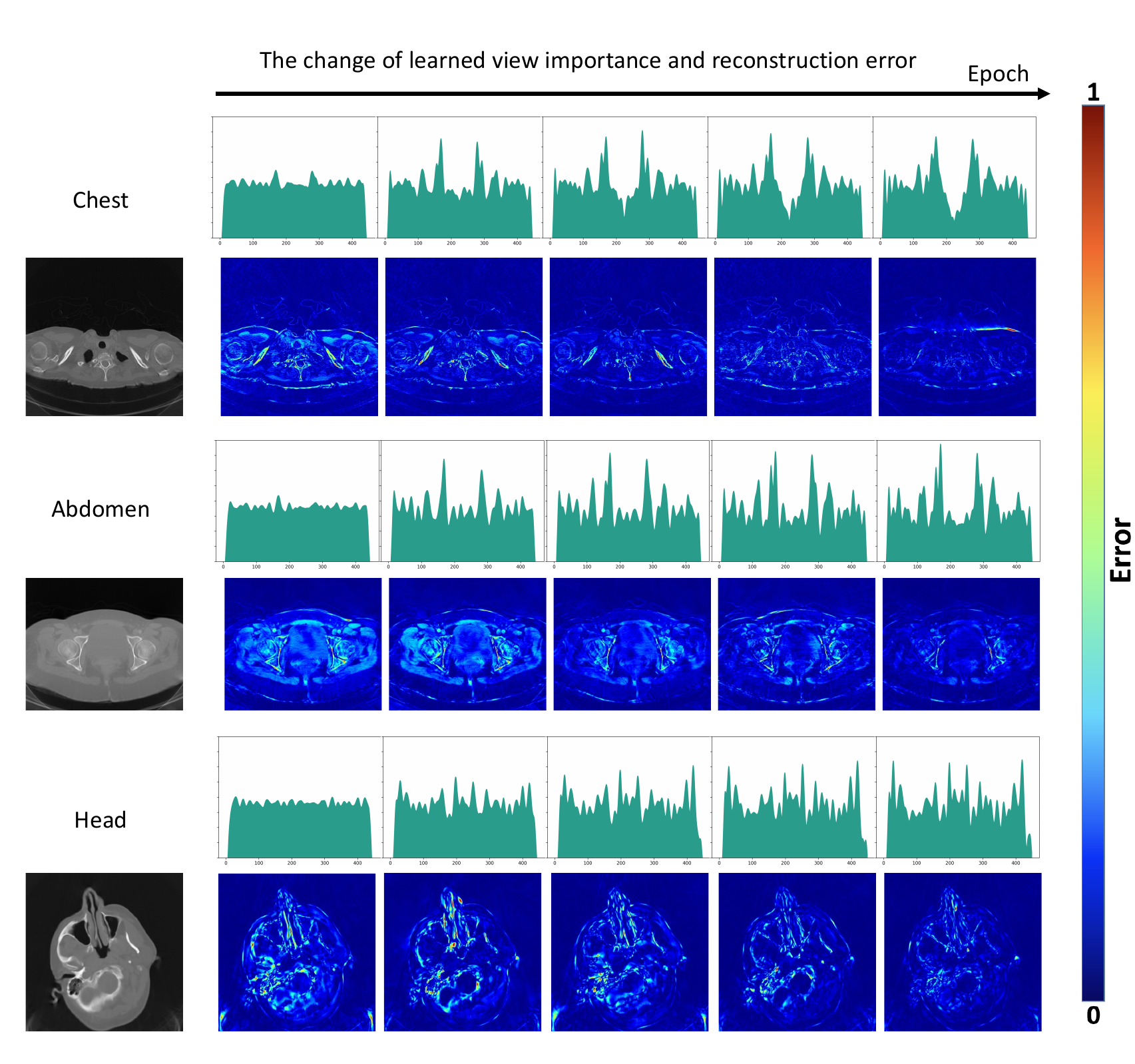}
	\end{center}
	\caption{The evolving trend of learned sampling strategies for different tasks during training demonstrates a reduction in the corresponding reconstruction errors as the strategies converge.}
	\label{views_epoch}
\end{figure}

\begin{table*}[t!]
\begin{center}
\caption{Quantitative results for ablation study. Best results are highlighted. $\downarrow$ ($\uparrow$) means the lower (higher) the better. * indicates a significant difference (p < 0.05) with baseline using Wilcoxon test.
 }
\label{t2}
\begin{tabular}{c||cc|cc}
\hline\hline
\rowcolor{gray!20} &\multicolumn{2}{c}{60 Views}\vline&\multicolumn{2}{c}{30 Views}\\
\rowcolor{gray!20} Method & PSNR($\uparrow$)&SSIM($\uparrow$)&PSNR($\uparrow$)&SSIM($\uparrow$)\\
\hline\hline
    
		Baseline & 38.62\scriptsize{$\pm$2.08}&0.940\scriptsize{$\pm$0.034}&36.51\scriptsize{$\pm$1.42} & 0.921\scriptsize{$\pm$0.042}
		\\
		Baseline+Multi-Task &39.18\scriptsize{$\pm$1.73}*& 0.942\scriptsize{$\pm$0.046}*&36.68\scriptsize{$\pm$2.11}*&0.926\scriptsize{$\pm$0.051}*\\
		Baseline+Multi-Task+Full-View Rep&  \cellcolor[HTML]{D7FFD7}39.48\scriptsize{$\pm$2.33}*&\cellcolor[HTML]{D7FFD7}0.943\scriptsize{$\pm$0.034}*&\cellcolor[HTML]{D7FFD7}36.90\scriptsize{$\pm$2.15}*& \cellcolor[HTML]{D7FFD7}0.927\scriptsize{$\pm$0.045}*\\
		

		\hline
		
	\end{tabular}
\end{center}
\end{table*}

\subsection{implementation details}
We implemented our framework in \textit{Python} with \textit{Pytorch} \cite{paszke2017automatic}. In task-specific sampling, we use 100 Gaussian center for \textit{Gaussian Mixture} distribution as the default experiment setup and the mixing coefficients $\pi_{k}$ are set equally for each center. We initialize the \textit{Gaussian Mixture} distribution uniformly and set $ \tau $ in gamble-max to 1. 
\begin{figure}[t]
\begin{center}
		\includegraphics[width=1\linewidth]{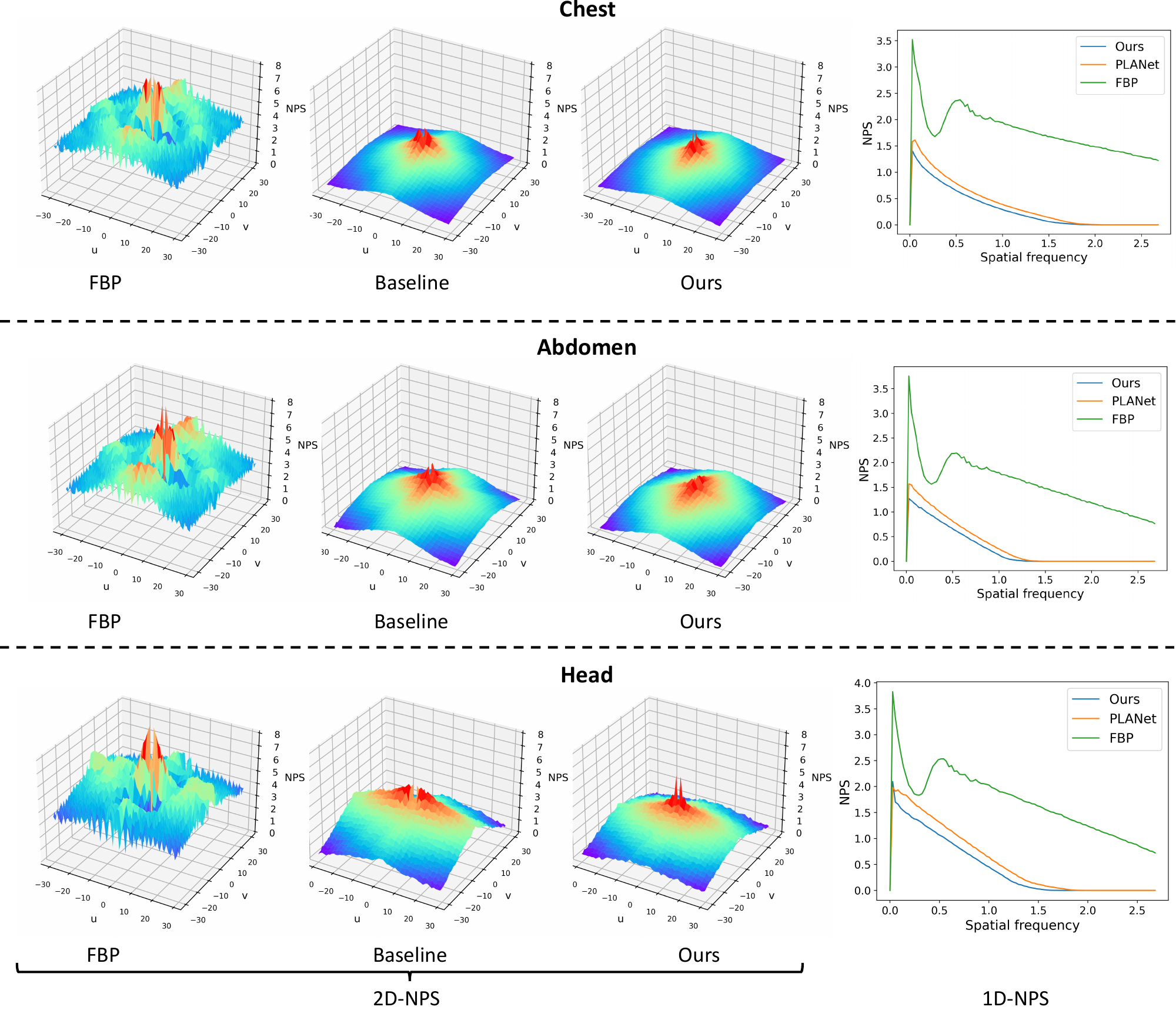}
	\end{center}
	\caption{Visualization of Noise Power Spectrum (NPS) for different reconstruction tasks. 2D-NPS represents the average results over all test images, while 1D-NPS is circularly averaged from the 2D-NPS.}
	\label{nps}
\end{figure}
For the downstream-task network, the choice of the backbone network depends on the task itself. Since the purpose of the experiments is to verify the effectiveness of task-specific sampling on downstream tasks, the principle is to choose a network that is as lightweight as possible. For COVID-19 positive/negative classification on the iCTCF dataset, we find that a lightweight ResNet-18 \cite{he2016deep} can achieve promising performance. For COVID-19 ten-day survival prediction on the Saber-Italy dataset, which is a more challenging task when using only CT images, we use DenseNet-121 \cite{huang2017densely} as the backbone. 

 \subsection{Results on Sole Reconstruction}
For the performance comparison, we selected six state-of-the-art methods: DDNet \cite{zhang2018sparse} is a representative image domain method that aims to remove artifacts and improve image quality from FBP-reconstructed images. HDNet \cite{hu2020hybrid} is a typical dual-domain method that interpolates sparse-view sinogram data and refines image quality using two CNNs separately. IradonMap \cite{he2020radon} is essentially a parameter-learnable Iradon transform coupled with an image domain CNN. RegFormer \cite{xia2023transformer} is a deep-unrolled reconstruction method combined with Swin-Transformer \cite{liu2021swin}. FreeSeed \cite{ma2023freeseed} is a frequency-band awarded method combined with a self-guided artifact refinement network. PLANet \cite{yang2022sparse} is a sampling strategy learning method which learns a universal sampling for all scanning types. All methods use 60 projection views to reconstruct images.

Quantitative results are shown in Table \ref{t1}. In the 60 views condition, FBP-reconstructed images suffer from severe artifacts, and all deep learning methods largely improve image quality. Comparing to uniform sampling reconstruction, the sampling strategy learning methods (PLANet and ours) gain better performance. By utilizing multi-task learning, the proposed method achieves the best performance in all three tasks. Visual results are shown in Fig. \ref{method_compare}. It demonstrates that task-specific sampling strategy can improve the performance on detail structures for each scanning type.

The violin plots for all three tasks are shown in Fig. \ref{violin}. The Friedman test is employed to detect significant differences in the performance of all reconstruction methods across the three tasks. In each task, we utilize the Wilcoxon test to verify whether the improvements are significant. The Friedman test shows that all reconstruction methods are significantly different. The Wilcoxon tests further demonstrate that our method has significant improvement over all other methods (with $p<0.05$).

The PLANet, which learns a universal sampling strategy for all tasks, exhibits more improvement in chest scanning. However, its performance drops in abdomen and head scanning. This suggests that the universal sampling strategy is biased towards chest data, which may hinder performance on other scan types. By employing task-specific sampling, our proposed approach addresses this issue and shows significant improvements across all tasks.

The learned sampling strategies are depicted in Fig. \ref{views}. We observe that scans for the chest and abdomen exhibit some similarity in view importance, but there also exist small differences between them. As for head scanning, the view importance differs significantly from both of them. This finding further verifies that the optimal sampling strategy varies according to the type of scans.
    

\subsection{Performance on Downstream-Task}

Two downstream-tasks are designed to verify the effectiveness of task-specific sampling. The first downstream-task is to conduct the experiment to classify Covid-19 positive/negative on iCTCF dataset. The second aims to predict the survive or death of scanned patients in ten days on saber dataset. The performances are evaluated in three metrics Accuracy(Acc), Sensitivity(Sens), and Specificity(Spec). We train classifiers with images that are reconstructed using different sampling strategies: 1) Full-view images (GT); 2) Sparse-view images using uniform sampling and reconstructed with FBP (FBP); 3) Sparse-view images using uniform sampling and reconstructed with deep learning network, which use the same network structure as the proposed method but trained with uniform sampling (Uniform Sampling); 4) Sparse-view images using learned task-specific sampling and reconstructed with proposed network (Task-Specific Sampling).

As the quantitative results in Table \ref{downstream} show, sparse-view CT images reconstructed using FBP result in a significant performance drop. This drop is particularly pronounced in the challenging task of predicting survival or death within ten days on the Saber dataset, where direct reconstruction with FBP fails to yield useful results. However, when reconstructed with a deep learning network (i.e., Uniform Sampling), classification accuracy improves in both downstream tasks. Furthermore, after learning the task-specific sampling for each task, the performance becomes comparable to that of full-view images. This indicates that learning sampling strategies can greatly benefit the clinical usage of sparse-view CT. Fig. \ref{downstream_f} presents a visualized case from the Saber Dataset. In this case, task-specific sampling reconstruction exhibits higher quality in the lesion area compared to uniform sampling, resulting in better performance in downstream tasks. Additionally, Fig. \ref{viewsd} illustrates the learned importance of views and selected imaging angles. We observe that the learned importance exhibits different trends for different tasks, and all tasks show improvement from the uniform sampling.

\subsection{Ablation Study}
To evaluate the effectiveness of multi-task learning, we compare the performance with the baseline \cite{yang2022sparse} using both 60 and 30 views. Additionally, we conduct a common multi-task framework following the approach in \cite{girdhar2022omnivore} to assess the effectiveness of full-view representation. As shown in Table \ref{t2}, both multi-task learning and full-view representation effectively improve performance. The results indicate that the improvements in image quality are more significant when using 30 views compared to 60 views, highlighting the importance of the task-specific sampling strategy, particularly when the number of views is limited.

Fig. \ref{views_epoch} illustrates the evolution of learned sampling strategies during training. Initially, we uniformly initialize the importance of view $P^{t}$ for all tasks, resulting in similar strategies for each task. However, as training progresses, these strategies evolve differently to better suit the characteristics of each task. Consequently, the reconstruction error is reduced as a result of the changing sampling strategies.

As the shape of the Noise Power Spectrum (NPS) reveals where the noise power is concentrated in frequency space, we compute the NPS across different tasks to evaluate the performance of task-specific sampling in the frequency domain. The NPS is calculated following the method proposed in \cite{boedeker2007application}, and the visualized results for three reconstruction tasks are presented in Fig. \ref{nps}, with the logarithm taken for visualization. The 2D-NPS for three methods (i.e., FBP, baseline method, and the proposed method) is initially plotted by averaging over all test images to depict the noise features in frequency space. For a clear comparison, we also circularly average the 2D-NPS into 1D-NPS. From Fig. \ref{nps}, we observe that noise exhibits different characteristics across various reconstruction tasks. With the task-specific sampling strategy, the proposed method demonstrates better performance than the baseline method in both the low-frequency and high-frequency areas.

\section{Conclusions}

In this study, we introduce a novel approach to address the challenge of severe artifacts in Sparse-View Computed Tomography (SVCT) by proposing a task-specific sampling strategy learning method. Our method leverages a multi-task learning framework to optimize imaging quality under sparse-view conditions for different CT scan tasks.
The primary cause of artifacts in SVCT is the loss of information due to insufficient data sampling. By optimizing the sampling strategy, we aim to enhance data quality from the source by selecting imaging views with higher importance. However, using a universal sampling strategy for all imaging tasks may not be optimal, as it may not effectively represent the data characteristics of each specific task. Therefore, our approach focuses on learning tailored sampling strategies that account for the differences between different types of tasks.
Experimental results showcase the effectiveness of our method in learning optimal sampling strategies for specific CT scan tasks, surpassing state-of-the-art methods across various datasets. Furthermore, experiments conducted on downstream tasks demonstrate the potential of our approach to improve the performance of clinically relevant tasks.
In conclusion, our study contributes a valuable advancement in SVCT reconstruction by introducing task-specific sampling strategy learning. This not only enhances imaging quality but also holds promise for improving the performance of clinical tasks. By addressing the inherent variability across different scanning tasks, our method offers a pathway towards more precise and efficient diagnostic imaging, ultimately benefiting patient care and outcomes in clinical settings.
 
%

\bibliographystyle{IEEEtran}
\bibliography{sample-base}

\vfill

\end{document}